\documentstyle[emulateapj,psfig]{article}
\lefthead{COLE ET AL.}
\righthead{UV POLARIMETRY OF THE LMC}

\def\cgssb{$\times$ 10$^{-8}$ erg s$^{-1}$ cm$^{-2}$ \AA$^{-1}$ Sr$^{-1}$ }
\def\surfb{mag arcsec$^{-2}$}
\def\hi{H {\small I}}
\def\ebv{E$_{B-V}$}

\begin{document}

% \begin{center}
% {\bf \today}
% \end{center}

\title{Ultraviolet Imaging Polarimetry of the Large Magellanic Cloud. I. Observations}

\author{Andrew A. Cole,\altaffilmark{1,4} Kenneth H. Nordsieck,\altaffilmark{1,2}
Steven J. Gibson,\altaffilmark{3} \& Walter M. Harris\altaffilmark{2}}
\altaffiltext{1}{Department of Astronomy, University of Wisconsin-Madison,
  475 North Charter St., 5534 Sterling Hall, Madison, WI, 53706; 
  {\it cole@astro.wisc.edu}}
\altaffiltext{2}{Space Astronomy Laboratory, University of Wisconsin-Madison,
  1150 University Avenue, Madison, WI, 53706; {\it khn@sal.wisc.edu}}
\altaffiltext{3}{Department of Physics \& Astronomy, University of Calgary, 
  2500 University Drive NW, Calgary, Alberta, T2N 1N4, Canada.}
\altaffiltext{4}{{\it Current address:} Department of Physics \& Astronomy,
  University of Massachusetts, Amherst, MA 01002--4525.}

\setcounter{footnote}{0}

\begin{abstract}
We have used the rocket-borne Wide-Field Imaging Survey Polarimeter (WISP) to 
image a $1\fdg5 \times 4\fdg8$ area of the western side of the
Large Magellanic Cloud (LMC) at a wavelength of $\lambda =
2150$ \AA$ $ and a resolution of 1$\arcmin \times 1\farcm5$.
These are the first wide-field
ultraviolet polarimetric images in astronomy.
We find the UV background light of the LMC to be linearly
polarized at levels ranging from our sensitivity limit of 4\% to 
as high as $\approx$40\%.  In general, the polarization in a pixel
increases as the flux decreases; the weighted mean value of
polarization across the WISP field is 12.6\% $\pm$2.3\%.  
The LMC's diffuse UV background, in uncrowded areas, rises from a minimum of 
5.6 $\pm$3.1 \cgssb (23.6 $\pm$0.5 \surfb) to
9.3 $\pm$1.1 \cgssb (23.1 $\pm$0.2 \surfb) in regions near the 
bright associations.   We use our polarization maps to investigate
the geometry of the interstellar medium in the LMC, and to 
search for evidence of a significant contribution of scattered
light from OB associations to the diffuse galactic light of the LMC.
Through a statistical analysis of our polarization map, we identify
9 regions of intense UV emission which may be giving rise to 
scattering halos in our image.
We find that starlight from the N 11 complex and 
the LH 15 association are the strongest contributors to the scattered
light component of the LMC's diffuse galactic light.  This region
of the northwestern LMC can be thought of as a kiloparsec-scale reflection nebula
in which OB stars illuminate distant dust grains, which scatter the
light into our sightline.  In contrast, the polarization map 
does not support the scattering of light from the large B2 complex
in the southern WISP field; this effect may be astrophysical, or it
may be the result of bias in our analysis.
\end{abstract}


\section{Introduction}

The presence of extended optical light halos around spiral galaxies
raises the question: is this light due to an extended stellar distribution,
or is it the result of scattered starlight from more central regions?
\cite{jur80} pointed out that if the extended clouds of neutral
hydrogen around late-type galaxies are associated with dust, the
clouds should appear as faint reflection nebulae.  In their investigation
of the diffuse ultraviolet light of the Galaxy, \cite{mur95} found a significant
contribution due to the scattering of starlight from OB stars by
interstellar dust. Photometric studies of spiral galaxies have provided
very suggestive evidence for UV scattering halos far from regions
of star-formation (\cite{ste82}, \cite{nef94}).  Signatures of 
dust-scattered starlight are spatial correlations with \hi, 
bluer colors than the general radiation field, and the presence
of linear polarization.  

The LMC, due to its proximity, vigorous star-forming activity, and
low level of obscuration by Galactic dust, is an ideal location
in which to investigate the connection between starlight and diffuse
light in a spiral galaxy.  The LMC is known to possess
a higher gas-to-dust ratio than the Galaxy 
(e.g., \cite{cla96}); while this would tend
to lessen the contribution of scattering to its diffuse galactic
light, the relatively high albedo of LMC dust (due to its depletion
in carbon relative to the Milky Way; \cite{pei92},
\cite{fit86}) should compensate somewhat.  

The Wide-Field Imaging Survey Polarimeter (WISP) instrument is 
designed to obtain polarimetric images of astronomical objects
in the ultraviolet in order to investigate the composition and
distribution of dust in astrophysical environments.
The ultraviolet is chosen because the emission from late-type 
stars, which are smoothly distributed and numerous, is suppressed.
At optical wavelengths, late-type stars contribute strongly to the
general radiation field of a galaxy and may mask the presence of
scattered light with their own radiation.  In the ultraviolet, OB
stars in spatially compact associations dominate the radiation field
(e.g., \cite{par98} for the LMC); far from these associations, scattered light at 
these wavelengths is expected to contribute strongly to the diffuse
radiation field.  In addition to the low stellar background, the 
UV is an ideal regime in which to make polarimetric investigations
because of high polarimetric efficiency
(\cite{nor93}).

If scattering is a significant contributor to the diffuse radiation
field, then its surface brightness and polarimetric properties can
be used to derive information regarding the distribution of dust
in the interstellar medium, as well as to place constraints on the
optical properties of the dust grains.

WISP has been flown three times to date; its targets have been 
the reflection nebulosity in the Pleiades star cluster
(\cite{gib95}; \cite{gib97}), Comet Hale-Bopp (\cite{har97}),
and the LMC.

In this paper, we present polarimetric images of a 
field along the western side of the LMC.  These are the first
wide-field ultraviolet polarimetric images ever obtained
in astronomy; ground-based imaging polarimetry in the 
{\it rest frame} UV of radio galaxies and QSOs at redshifts
$z$ $\sim$ 1 has been analyzed by, e.g., \cite{cim94}, \cite{dis93},
and references therein.  \cite{cap95} used the Hubble Space
Telescope's Faint Object Camera to obtain a UV polarimetric
image of the galaxy NGC 1068 with an 11$\arcsec$ $ $ field of view.

We describe the WISP
instrument in \S\ref{inst}; in \S\ref{data} we provide the flight information 
and observation sequence, and describe the procedures we have
adopted for reduction and calibration of the images. 
\S\ref{surf} gives the results of our surface photometry
of the diffuse galactic light of the LMC.
In \S\ref{map} we present our polarization map and our analysis which
reveals the scattering halos around some of the prominent
concentrations of young stars.  In \S\ref{anal} we analyze the morphology
of the UV image and polarization map and how it fits in with
what is known about the structure and content of the ISM in 
this region.  We discuss in \S\ref{dis} the origin of the diffuse UV
radiation field in the LMC and the scattering geometry within
the LMC disk; our results are summarized in \S\ref{sum}.

\section{The Instrument }
\label{inst}

The Wide-field Imaging Survey Polarimeter (WISP) is a suborbital
rocket payload based around a 20 cm, F/1.8, off-axis, all-reflective
Schmidt telescope.  Nordsieck {\it et al.} (1993) provides a detailed
description of the payload optics.  The Schmidt corrector mirror removes
the spherical aberration of the primary, providing a 1$\fdg$5 $\times$
4$\fdg$8 degree field of view with resolution well-suited to the 
15 arcsecond pixel scale of the CCD detector.  The WISP detector for
this flight was a thick, front-illuminated Reticon 400 $\times$ 1200
pixel CCD, phosphor overcoated to provide UV sensitivity.  The quantum
efficiency of the detector in the vacuum ultraviolet is approximately
15\% and the readnoise is 10 e$^-$.  The CCD was cooled to $-$65
$\arcdeg$C by a thermo-electric cooler which dissipates its heat into
a copper block which is precooled to $-$40 $\arcdeg$C on the ground.
WISP has two broadband filters, at 1640 $\pm$250 \AA\ and 2150 $\pm$300
\AA, each comprised of aluminum/MgF$_2$ thin films on MgF$_2$ and
silica substrates.

The polarimetric analyzer is a large flat mirror mounted between the
Schmidt corrector and primary, coated with ZrO$_2$ (a high-index dielectric),
and illuminated at its Brewster angle.  The polarimetric modulator is a 
rotatable 20-cm halfwave plate mounted in the telescope aperture.  The
waveplate is constructed from four panes of oriented CaF$_2$ crystal.
The crystal is made to be birefringent by applying stress to one side using
a programmable actuator.  The entire waveplate assembly is then rotated
to provide polarimetric modulation.  Four images provide the linear 
polarization Stokes parameters: ``Q$^+$'' is obtained with zero waveplate 
pressure; ``Q$^-$'', ``U$^-$'', and ``U$^+$'' with $\frac{1}{2}$ wave
retardation, rotated 45$\arcdeg$, 67$\fdg$5, and 22$\fdg$5 relative to
the Brewster mirror axis, respectively.  Because the modulator is the 
first element in the WISP optical path, instrumental polarization from
subsequent optical elements is eliminated.  The position angle zero-point
was fixed by optical alignment and mechanical tolerances to better than 
$\pm$1$\arcdeg$.

Polarimetric calibration was performed in August, 1995 using the 40 cm
vacuum collimator at the Space Astronomy Lab, University of Wisconsin.
100\% linearly polarized continuum ultraviolet light was created using
a D$_2$ hollow cathode lamp illuminating the focal plane aperture of the
collimator, immediately followed by an MgF$_2$ Brewster-angle polarizer.
The polarimetric efficiency of the analyzer is estimated to be better
than 98\%, based on crossing two analyzers in visible light, and on 
analysis of the effect of small changes in the Brewster angle in the 
vacuum ultraviolet.  Collimated light from this device illuminated the
center and extrema of the WISP field by maneuvering the payload on a 
remotely controlled tilt table.  The efficiency of the WISP analyzer was
evaluated by rotating the calibration polarizer with the waveplate off;
it is 94\% in the 2150 \AA\ filter.  The optimal stress for halfwave 
retardation was evaluated by fixing the calibration polarizer angle
and WISP waveplate angle, while varying the waveplate stress.  The 
waveplate efficiency was then evaluated by modeling the polarimetric
modulation with waveplate angle at the optimal stress, and found to
be 79\% in the 2150 \AA\ filter.  The net polarimetric efficiency in
this filter is therefore $\approx$74\%; a
figure of 70\% was used during reductions.

\section{Data}
\label{data}

\subsection{Flight and Observations}

WISP was flown from Woomera, Australia on 17 November, 1995;
the flight was a complete success.
For the LMC flight the 15$\arcsec$ CCD pixels were binned to 4 columns
by 6 rows, resulting in an image scale of 1$\arcmin$ 
$\times$ 1$\farcm$5 per pixel.  The 2150 \AA$ $ filter was used
to take a polarimetric sequence of 4 80-second exposures, in the order
Q$^-$, Q$^+$, U$^-$, U$^+$.  The science exposures began at an
altitude of 230 km, continued through the apogee of 328 km, and 
ended at 150 km.  The planned pointing placed the long axis of
the CCD 9$\arcdeg$ from north-south, with the field centered at
$\alpha$ = 4$\mathrm ^h$ 58$\mathrm ^m$ 45$\mathrm ^s$, 
$\delta$ = $-$67$\arcdeg$ 55$\arcmin$ (B1950.0).  The pointing 
drifted by less than 5$\arcmin$ between the first and last
exposures. 

\subsection{Reduction}

We reduced the images within IRAF\footnote{IRAF is distributed by the National Optical
Astronomical Observatories, which are operated by the
Association of Universities for Research in Astronomy, Inc.,
under cooperative agreement with the National Science Foundation.},
removing the bias and dark count in the usual way.  
During the flight, a sequence of 8 bias frames and 
3 40-second dark frames were taken for calibration.
Additionally,
a sequence of 12 dark exposures taken in the laboratory were combined
to produce a superdark frame.  The lab darks confirmed the linearity
of the CCD, and enabled us to better characterize the small-scale
variations in pixel response when subtracting the dark frames. 

The rocket had begun its descent by the time of the 
last exposure, and a significant additional sky 
background was present in the U$^+$ image.  We corrected
for this by requiring that the sum of the Q images 
be equal to the sum of the U images, subtratcting 
the difference in sky background from the U$^+$ image.
The correction amounted to $\sim$16 counts, some 7.7\%
of the total sky background.

We removed cosmic ray contamination in a two-step process.
First, we used the IRAF task {\tt cosmicrays} to find and
remove pixels which deviated from the median of their neighbors 
by greater than a factor of ten.   These regions were replaced
by the mean of their neighbors.  The remaining cosmic rays
were found during a trial combination of the Stokes images,
which rejected pxiels that were more than 10$\sigma$ deviant
from the mean across all four images.  Deviant pixels were
masked off and excluded from further analysis.  10 bright,
compact areas of emission, well-distributed across
the field of view, were used to align and trim the images.
The reduced images were combined to create intensity,
Stokes Q, and U images:

\begin{eqnarray}
\mathrm I  =  Q^+ + Q^- + U^+ + U^-,\\
\mathrm Q  =  Q^+ - Q^-,\\
\mathrm U  =  U^+ - U^-.
\end{eqnarray}

% \noindent Our reduced image is shown in Figure \ref{iqu}.

\begin{figure*}
\centerline{\hbox{\psfig{figure=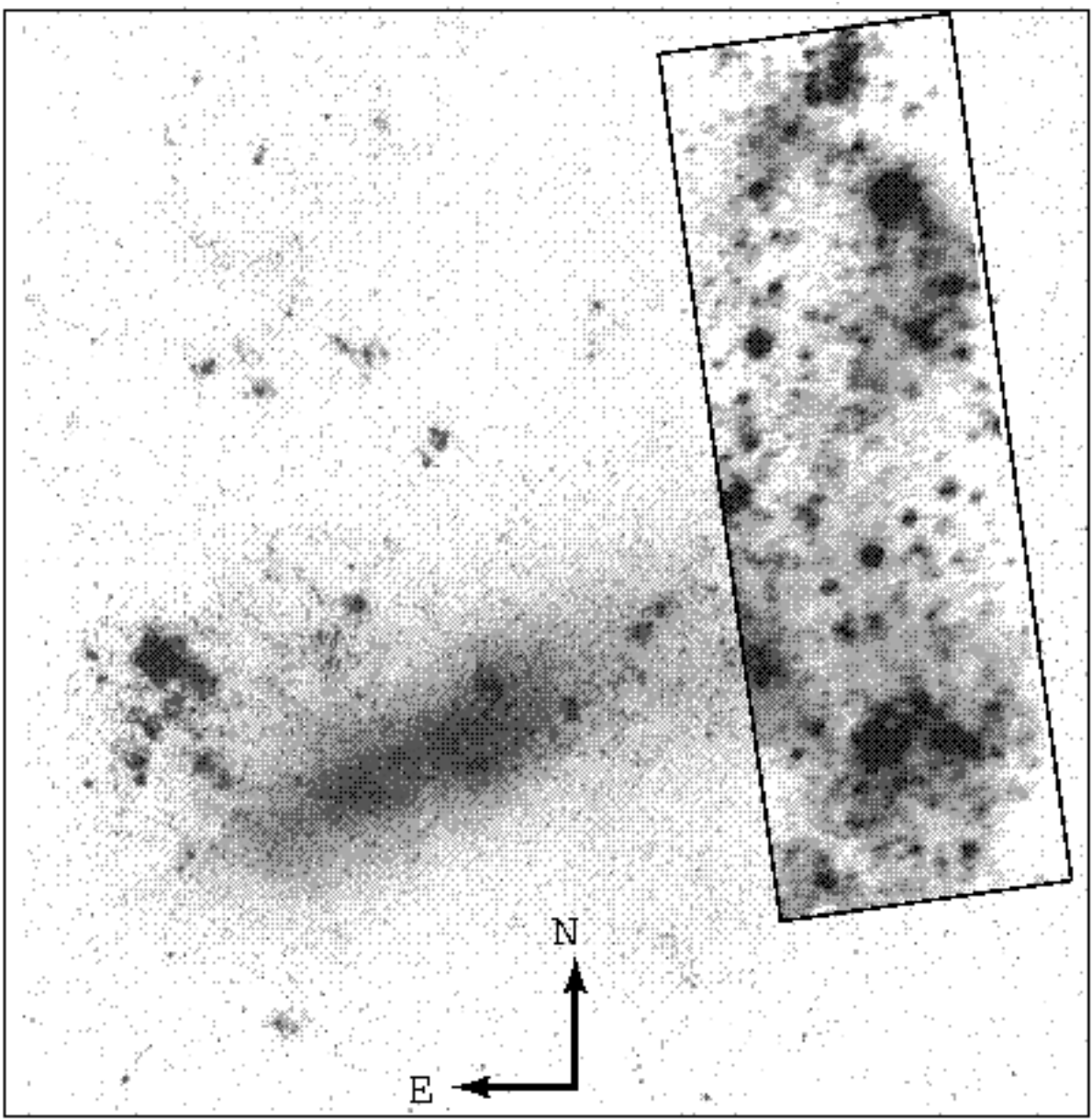}}}
\caption{The location of the WISP field within
the LMC.  The underlying optical image 
($\lambda$ $\approx$ 6600 \AA) was taken by Karl Henize;
see Sandage (1961).  The inset WISP field measures 1$\fdg$5 $\times$
4$\fdg$8 and is centered at $\alpha$ $\approx$ 5 hours, $\delta$
$\approx$ $-$69$\arcdeg$; the center of the LMC is at $\alpha$ = 
5$^{\mathrm{h}}$ 24$^{\mathrm{m}}$, $\delta$ = $-$69$\arcdeg$
48$\arcmin$.
\label{lmc}}
\end{figure*}

\begin{figure*}
\centerline{\hbox{\psfig{figure=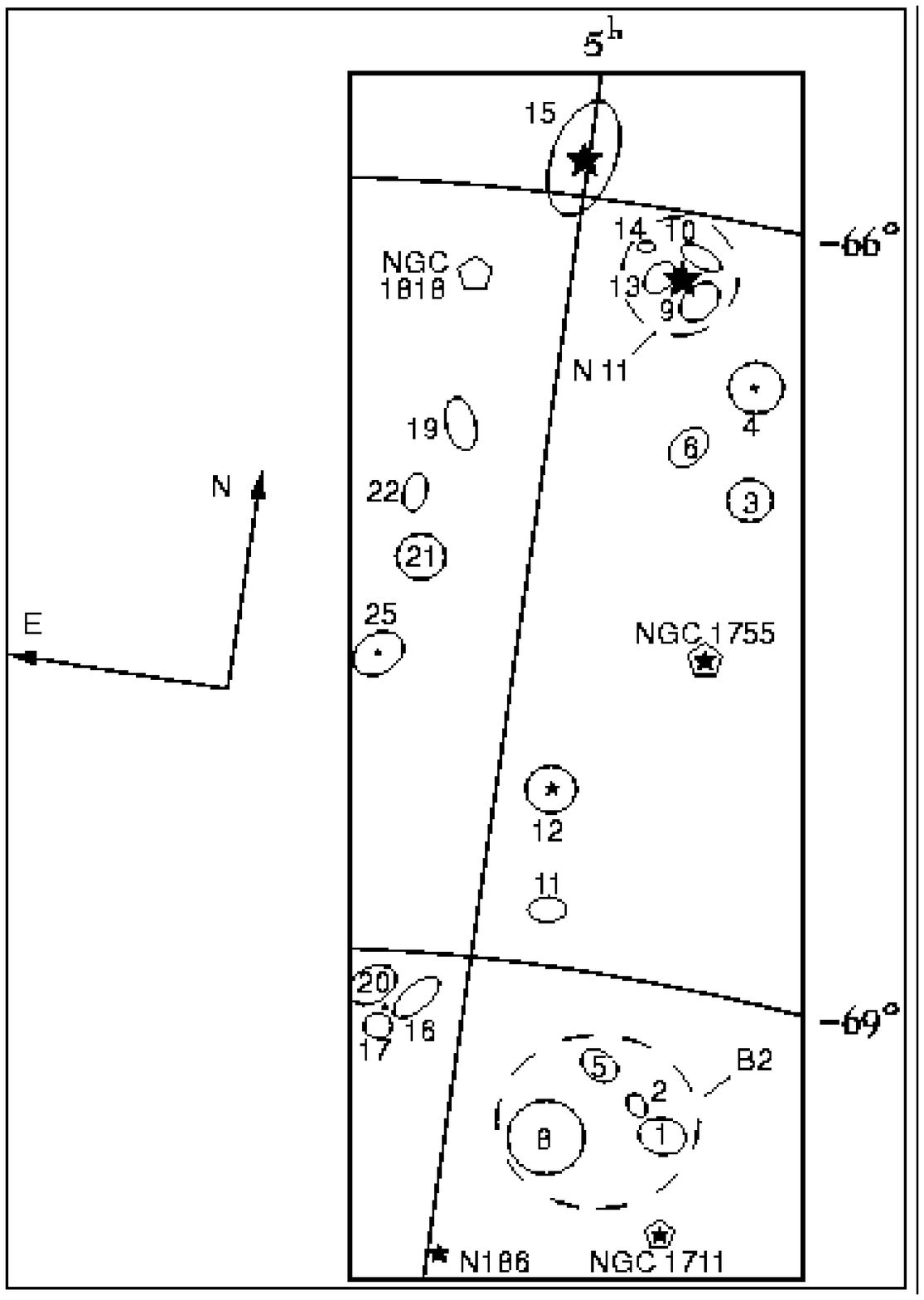}}}
\caption{Prominent features 
in the WISP field, after Lucke \&
Hodge (1970).  Numbered ellipses represent the positions of OB associations
from Lucke \& Hodge; the dashed ellipses N 11 and B2 refer to the enclosed
OB associations and emission nebulae.  Massive open clusters are denoted
by pentagons and labelled with their NGC number.  Sources of scattered
light from Table 2 are marked with stars; large stars indicate
stronger detections.
\label{toon}}
\end{figure*}

The intensity image was compared against the catalog of 
\cite{luc70}, the atlas of \cite{hod67}, the UV images
of \cite{smi87} (hereinafter SCH), and the catalog of \cite{san70} in order
to identify emission sources.  We associated 21 bright 
regions with the OB associations from \cite{luc70}.  
Other features include young open clusters, and in some
uncrowded regions we detect individual, early-type
supergiants (\cite{san70}).  We placed the image on
a world coordinate system using some of these identifications
and found the long axis of the image to be aligned
8$\arcdeg$ $ $ east of north.  The greyscale intensity
image is overplotted on an optical image of the LMC
in Figure \ref{lmc}. 
A schematic figure depicting
some of the emission sources is shown in Figure \ref{toon}.

\subsection{Photometric Calibration}

SCH presented photometry at 1500 \AA$ $ and 
1900 \AA$ $ for each association in the \cite{luc70}
catalog.  We used their results to obtain a photometric
calibration of our data, using the equation

\begin{equation}
\label{schcor}
m^{syn}_{21} = m_{19} - (m_{19} - m_{21})_0 - E_{19-21},
\end{equation}

\noindent where $m_{19}$ refers to the measured 1900 \AA$ $
magnitude from SCH, $(m_{19} - m_{21})_0$ refers to the
intrinsic 1900$-$2150 \AA$ $ color of an association, and
E$_{19-21}$ denotes the amount of selective
extinction between the two wavelengths. 
$m^{syn}_{21}$ refers to the synthetic 2150 \AA$ $ 
magnitude\footnote{We adopt throughout the definition
$m_{\lambda}$ = $-$2.5~log~($\mathrm F _{\lambda}$) $-$ 21.1.}
of each association, calculated from the quantities on
the right-hand side of Eq. \ref{schcor}.

For a flat spectrum source
with zero reddening, F$_{\lambda}$ = constant, and
Eq. \ref{schcor} becomes $m^{syn}_{21}$ = $m_{19}$.
The two corrective terms arise because the complex
spectral energy distribution of the associations
and the wavelength-dependent interstellar extinction
distinguish the two filter bandpasses from each other.

The list of calibrating associations is
given in Table 1, which includes the magnitude at 
1900 \AA$ $ and 1500$-$1900 \AA$ $ color from SCH, the
aperture used by SCH, and the \ebv$ $ for each
association from \cite{luc74}.

We dereddened the photometry of SCH, using the value
of \ebv$ $ from Table 1 and assuming that the maximum
contribution of the Galactic foreground was 0.07 mag.  
\cite{luc74} does not give \ebv$ $ for the associations
LH 10, 11, 13, or 14 and so we have estimated the reddening
to be 0.07.
We determined A$_{1500}$/\ebv$ $ and A$_{1900}$/\ebv$ $
by convolving the  extinction curves of \cite{car89} and \cite{fit86},
assuming R$_V$ = 3.1, with the source spectrum of a
T$\mathrm _{eff}$ = 30,000 K star (\cite{kur91}) and
the filter transmission curves published by SCH.  The
resulting reddening corrections were:

\begin{eqnarray}
m_{15,0} = m_{15} - 8.82\, E_{B-V}^{LMC} - 8.23\, E_{B-V}^{MW}
\\m_{19,0} = m_{19} - 8.28\, E_{B-V}^{LMC} - 8.21\, E_{B-V}^{MW}.
\end{eqnarray}

We then used the calculated $(m_{15} -  m_{19})_0$ color, along
with stellar atmospheres from \cite{kur91} and the filter transmission
curves of SCH and the WISP 2150 \AA$ $ filter, to predict the 
$(m_{19} - m_{21})_0$ color of each association.  For OB stars,
we adopted a linear relation between the two colors:

\begin{equation}
\label{sed}
(m_{19} - m_{21})_0 = 0.57\, (m_{15} -  m_{19})_0.
\end{equation}

\noindent This relation becomes invalid for sources cooler than
T$\mathrm _{eff}$ = 12,000 K.  The intrinsic color of the sources
suffers additional reddening due to the well-known extinction 
bump at 2175 \AA.  Using the LMC
extinction curve of \cite{fit86} and the Milky Way curve
of \cite{car89}, we calculate 

\begin{equation}
\label{red}
E_{19-21} = - E_{B-V};
\end{equation}

\noindent note that because of the 2175 \AA$ $ bump, 
$(m_{19} - m_{21})$ becomes {\it bluer} with increasing
extinction\footnote{The adoption of a mean UV extinction
curve is fraught with peril, as well-documented by
\cite{fit98}; variation in the extinction curve
from sightline to sightline may contribute an additional
20\% systematic uncertainty in photometry.}

Substituting Equations \ref{sed} and \ref{red} into Eq. \ref{schcor},
we obtain

\begin{equation}
m^{syn}_{21} = m_{19} - 0.57\, (m_{15} -  m_{19})_0 + E_{B-V}.
\end{equation}

\noindent The results for each association are given in 
column 7 of Table 1.

We photometered the associations in the WISP image, 
attempting to match the apertures to those used by 
SCH.  SCH do not give the precise locations of their
apertures, but they provide the area of each region;
we matched the areas using rectangular apertures in 
the IRAF task {\tt polyphot}.  The instrumental 
magnitudes are given in column 6 of Table 1.

We rejected the association LH 20 because it fell along
the edge of the field; there was some uncertainty in
the position of LH 14.  We calibrated our photometry
using a least-squares fit to $m^{syn}_{21}$; allowing 
for a nonlinearity as well as zeropoint offset, we 
found $m^{syn}_{21}$ = (1.05 $\pm$0.10)\, $m^{inst}_{21}$ $-$
(9.33 $\pm$0.24), with an rms scatter of 0.26 mag.
However, this fit was only marginally better than one
in which we enforced linearity, and which we decided 
to adopt:

\begin{equation}
\label{cal}
m^{syn}_{21} = m^{inst}_{21} - (8.53 \pm0.28),
\end{equation}

\noindent with an rms scatter of 0.29 mag.  For comparison
to $m^{syn}_{21}$, our calibrated photometry for the 
20 measured associations is given in column 8 of Table 1.
LH 14 was excluded from the fit because of the uncertainty in
its identification; it is faint, small, and subject to uncertain
sky subtraction due its close proximity to the extremely bright
associations LH 9 and 10.  

\subsection{Background Subtraction}

UV surface brightness measurements are affected by three major
sources of background contamination (\cite{mur95}): zodiacal light, 
Galactic background light, and the extragalactic background.
In order to measure the diffuse galactic light of the LMC, we 
must remove these background sky levels.  \cite{mau80} surveyed
the entire sky at 2200 \AA$ $ with 3$\arcdeg$ resolution, and 
measured a mean background sky level (Galactic plus extragalactic)
of 0.9 $\pm$0.2 \cgssb 
for the Galactic latitude of the western LMC ($-$35$\arcdeg$
$<$ b $<$ $-$38$\arcdeg$).  They reported a zodiacal light level
at 2200 \AA$ $ of 0.10 $\pm$0.02 \cgssb for the LMC's ecliptic
latitude.  We therefore adopt a total background level of 
1.0 $\pm$0.2 \cgssb (25.4 $\pm$0.3 \surfb) across the WISP field.
This value has been subtracted from each of our surface brightness
measures of the LMC's diffuse galactic light.

\section{Surface Photometry}
\label{surf}

\begin{figure*}
\centerline{\hbox{\psfig{figure=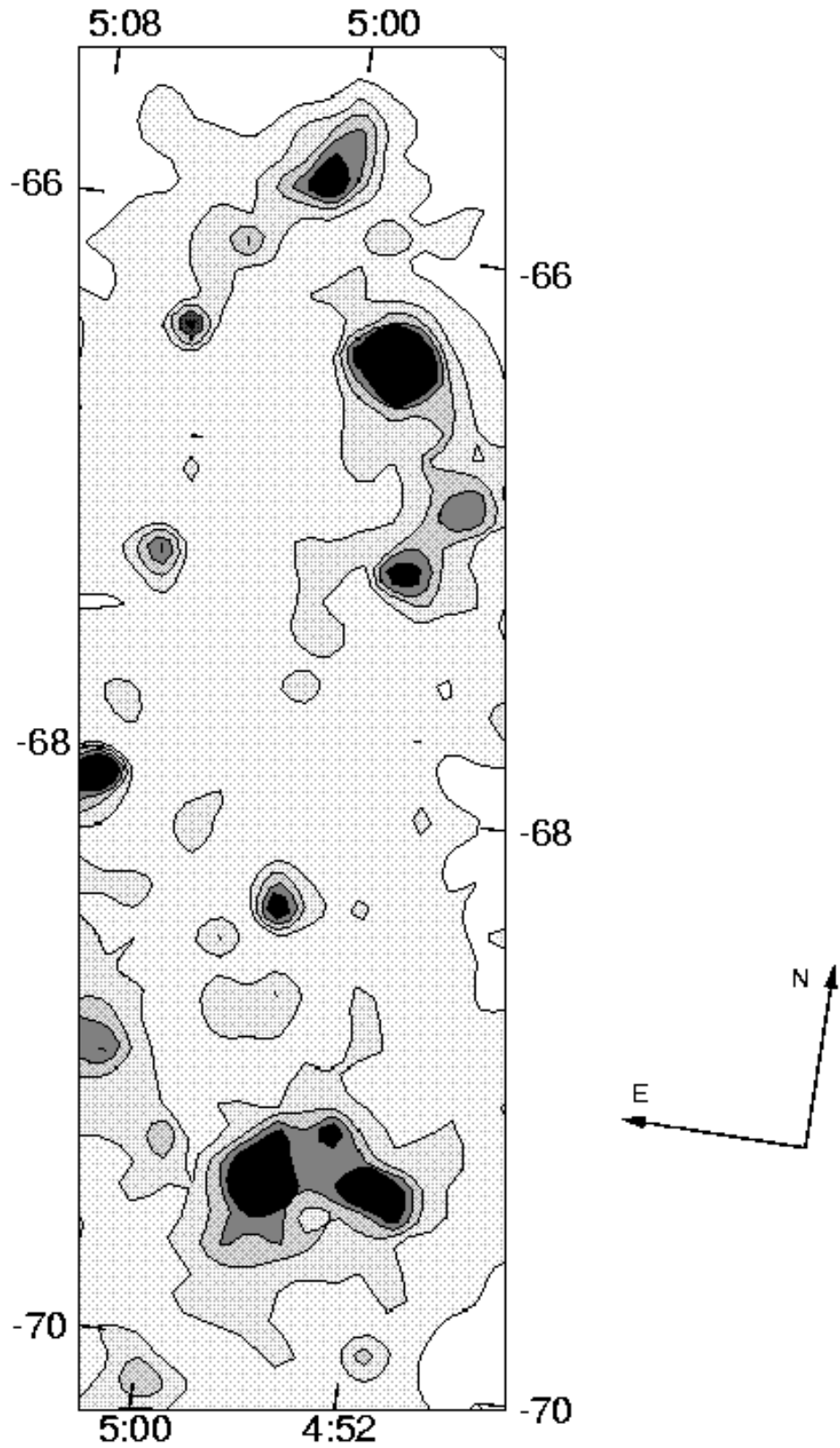}}}
\caption{Surface brightness plot of the WISP field;
the image has been smoothed to 0$\fdg$1 resolution.  North is up
and East is to the left.  Contour lines are drawn at
6, 11, 16, 21, 26 \cgssb (23.5, 22.9, 22.5, 22.2, 21.9 \surfb).
\label{cont}}
\end{figure*}

A contour plot of the LMC's surface brightness at 2150 \AA$ $ is 
shown in Figure \ref{cont}, at a resolution of 0$\fdg$1.
The diffuse UV background, which we
attempted to measure in uncrowded areas, rises from a minimum of
5.6 $\pm$3.1 \cgssb (23.6 $\pm$0.5 \surfb) to
9.3 $\pm$1.1 \cgssb (23.0 $\pm$0.2 \surfb) in regions near the
bright associations.  The lowest areas occur where the stellar 
density is low, at the northern edge of the image, along the 
western edge of the image near $-$68$\arcdeg$, and in the far
southwestern corner.
Uncrowded regions of high surface brightness include the region 
between N 11 and LH 19, and the area north of NGC 1755.  
Regions above the 11 \cgssb contour line are highly resolved
in the SCH images into sheets of stars; Fig. \ref{cont} shows
that these high surface brightness regions are clustered around
OB associations.

The eastern edge of the image, at roughly $-$69$\arcdeg$, 
includes the extreme western end of the LMC bar.  The bar
is prominent in optical images of the galaxy, less so in
UV images, e.g., SCH.  The stellar density is far higher
here than in the northern half of the WISP field; despite
the suppression of late-type stars by the UV filter, the
densely populated bar contributes to the high UV surface
brightness.  As will be discussed in \S\ref{anal}, this
will adversely affect our ability to detect the effects
of dust scattering.

\cite{mau80} presented a surface brightness map of the LMC at 2200 \AA.
Agreement between the WISP surface photometry and the \cite{mau80}
results is surprisingly good, considering the disparity in angular
resolution and detector contstruction, as well as the uncertainties
in photometric calibration.  Due to resolution effects, Figure 
\ref{cont} contains more extreme values, both high and low, than
the \cite{mau80} map; the effect was lessened, although not eliminated,
by smoothing the WISP image to 1$\arcdeg$ resolution.
The difference is greatest for the degree-scale arcs seen in the
WISP image north of $\approx$ $-$67$\arcdeg$ 30$\arcmin$: they are 
blended out by the 3$\arcdeg$ resolution in the earlier work.
The arc containing N 11 is visible in the \cite{mau80} map
as a small perturbation to the 11 \cgssb contour line.  In the
southern WISP field, the light is more smoothly distributed,
and the 16 \cgssb contour in Fig. \ref{cont} matches the location
of the similar isophote in \cite{mau80}.

\section{Ultraviolet Polarization Map}
\label{map}

We mapped the degree of linear polarization, P, and position
angle, $\theta$ for the WISP field using the I, Q, and U images.
Because the direct light of stars contributes highly
to the flux but is largely unpolarized, their presence
tends to bias the estimates of the polarization of the
diffuse galactic light to low values.  To remove them
from the polarization map, we cleaned the images of 
pixels that deviated by more than 5$\sigma$ above the
local sky value in all three of the images.  The star mask
was applied to the individual images and masked pixels
were subsequently ignored.

In order to increase the signal-to-noise, we binned the
image into 36 arcmin$^2$ pixels (4 $\times$ 6 pixels).
For each pixel we calculated the degree of linear polarization:

\begin{equation}
\mathrm P\; =\; \left( \frac{(Q^2\; +\; U^2)}{I^2} \right) ^{0.5},
\end{equation}

\noindent and the position angle, measured clockwise from
the positive $y$-axis (roughly north):

\begin{equation}
\theta \; =\; 0.5\; \arctan \left( \frac{U}{Q} \right) .
\end{equation}

We assumed Poisson statistics for the calculation of 
$\sigma _{\mathrm P}$ and $\sigma _{\theta}$.  Because
P is the sum in quadrature of Q and U, the expectation
value of P is biased such that for constant Q and U,
the most probable measured value of P increases with
increasing $\sigma _{\mathrm Q}$ and $\sigma _{\mathrm U}$.
Because our data suffer from poor signal-to-noise levels,
we have removed this bias using the formula of \cite{nor76}.
By this formula, a 2.5$\sigma$ detection is biased so
that P$_{obs}$/P$_{true}$ $\approx$ 1.1.  We reject 
as non-detections pixels in which 
P $<$ 2.5$\sigma _{\mathrm P}$.  This restriction 
set our detection limit at 4\% polarization.

\begin{figure*}
\centerline{\hbox{\psfig{figure=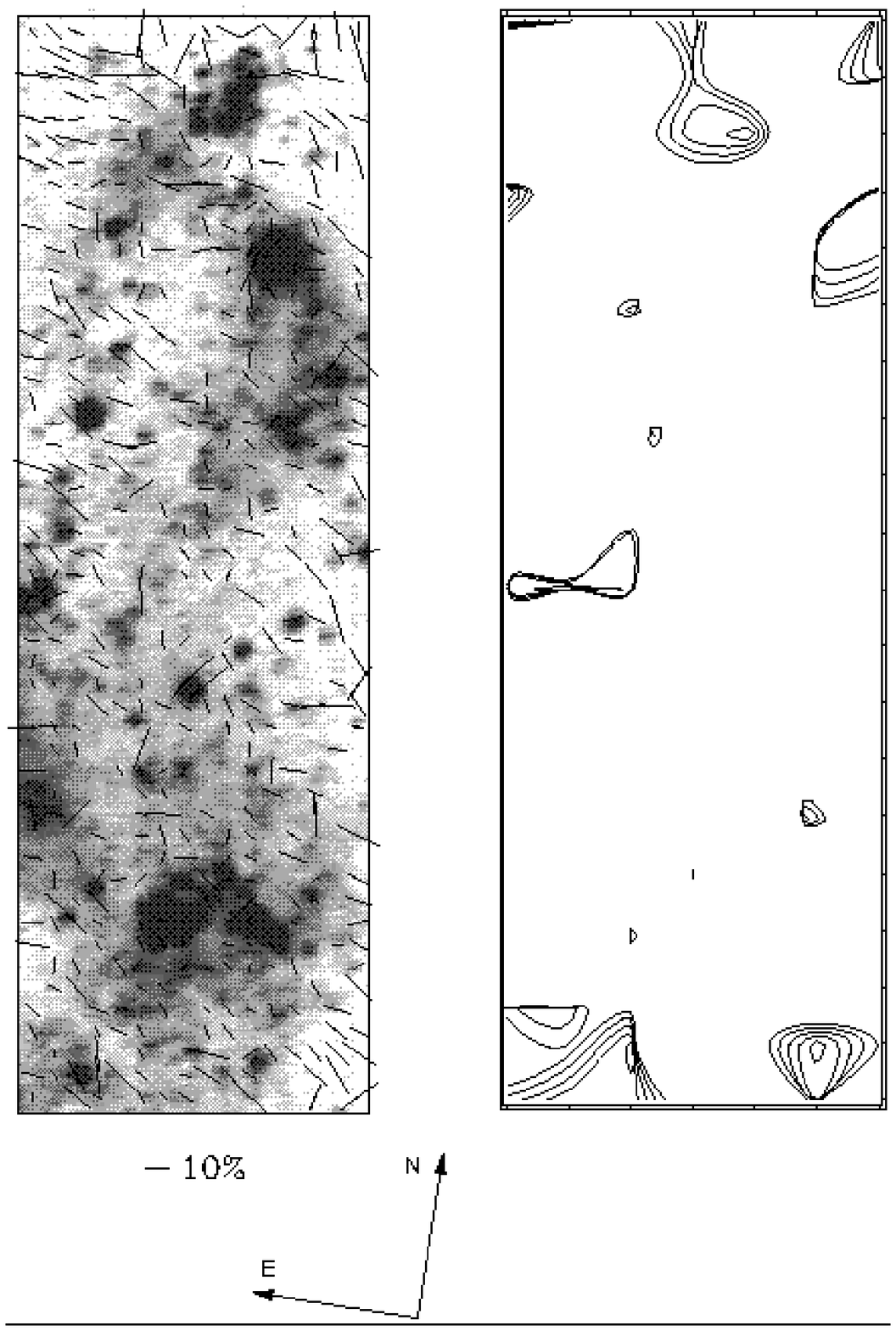}}}
\caption{{\sl a)} Ultraviolet polarization map
of the WISP field, binned to a resolution of
6 $\arcmin$ per pixel.  North is up and East is
to the left.  Our map becomes incomplete for
polarizations below 10\%, and our sensitivity
limit is 4\%.  Hints of centro-symmetric scattering
patterns can be seen.  {\sl b)} A smoothed contour
map of $\bar{\chi ^2}$, where the $\bar{\chi ^2}$
value of a pixel represents the likelihood that the
polarized light in neighboring pixels consists of
singly-scattered photons originating in the pixel
of interest.  Contours are plotted at $\bar{\chi ^2}$
= 1,2,3,4,5.
\label{pols}}
\end{figure*}

The final polarization map is shown in Figure \ref{pols}a.
Polarizations as high as 40\% are observed in some
pixels, although their absolute flux level is
small.  Most of the values lie between 5\% and 20\%,
with an error-weighted mean P = 12.6\% $\pm$2.3\%.
If we relax the restriction on detections below
2.5$\sigma$, the mean P = 4.4\% $\pm$7.8\%.  

The maximum predicted polarization from a commonly
used model of the interstellar dust mixture
(Mathis, Rumpl, \& Nordsieck 1977) is $p_{max}$
= 0.3--0.4 (White 1979).  We can estimate the
fractional contribution of scattered light to
the total UV surface brightness in the WISP field
by comparing our mean P to $p_{max}$.  This comparison
suggests that 10--15\% of the UV surface brightness
is due to polarized scattered light.  However, 
polarization as high as $p_{max}$ occurs only for
scattering at a 90$\arcdeg$ angle, and decreases
for smaller and larger scattering angles.  Because
of the forward-throwing nature of dust grains in the
ultraviolet, the mean scattering angle is expected
to be much smaller than 90$\arcdeg$ for our data.
The polarization of a scattered photon decreases
asymptotically to zero as the scattering angle
approaches zero (e.g., Scarrott {\it et al.} 1990).
We therefore predict that much more than 10--15\% 
of the measured UV surface brightness is contributed
by scattered photons.

For the case of scattering from a point source, the 
polarization vectors form a characteristic centro-symmetric
pattern around the illuminating source.  The typical error
in our position angles is $\pm$25$\arcdeg$; this noise
distorts the centro-symmetric signature of scattering
halos around bright sources, if such features are present.
Although our map is noisy, the high mean level of P is indicative 
of a significant scattered light component to the general
UV surface brightness.

To identify the sources most likely responsible for illuminating
the interstellar dust, we performed a statistical analysis
designed to evaluate the degree of centro-symmetry in the
polarization map.   We stepped through the polarization map,
pixel-by-pixel, and for each pixel, we examined the distribution
of polarization vectors in the encompassing 7 $\times$ 7 pixel 
(42$\arcmin$ $\times$ 42$\arcmin$) region.
For each region, we computed a $\chi ^2$ value based on the
distribution of ($\theta_{obs} - \theta_{pred}$), compared
to the observed position angle errors $\sigma_{\theta}$.
We thereby created a map of $\chi ^2$ values; the $\chi ^2$ 
of a given pixel is a measure of the likelihood that the 
measured polarizations in neightboring pixels have arisen
due to scattered photons originating in the pixel of interest.

% For each
% pixel, we used a $\chi ^2$ test to compare the observed
% polarization position angles in the surrounding 7 $\times$ 7
% pixel region to the position angles that would be observed,
% given point source illumination from the central pixel of 
% the region.  By comparing the distribution of ($\theta_{obs}
% - \theta_{pred}$) to the observed errors $\sigma_{\theta}$, 
% we were able to determine the probability, for each pixel
% in the image, that the pixel in question was the sole 
% contributor to the nearby polarized flux.  

The mean value of $\chi ^2$ was 8.3 $\times$ 10$^3$; several
minima, with $\chi ^2$ of order 1--10, were present.  
To minimize the impact false (noise) minima, we averaged over
6$\arcmin$ regions to produce our final $\bar{\chi ^2}$ map
(Figure \ref{pols}b).  We have highlighted the regions of 
low $\bar{\chi ^2}$ by plotting Figure \ref{pols}b using 
contours at 1,2,3,4, and 5.  The smoothing procedure has 
failed to eliminate the worst of the false minima, which
are typically shallow and spatially compact, or adjoin an
edge of the image.  Broad but shallow minima which correspond
to sources such as LH 4 or NGC 1755 (see Figure \ref{toon}, 
Table 2) have been obscured by the smoothing procedure and
do not appear in Figure \ref{pols}b.

We correlated the $\bar{\chi ^2}$ and intensity images
in order to identify the birthplaces of scattered photons.
During this process, we discovered that the locations of 
the $\bar{\chi ^2}$ minima and the centroids of the OB
associations were, in general, offset from each other.
This behavior is expected for light scattered from an
extended dust layer, inclined to the line of sight.
For the N11 complex, which is the brightest region in 
the WISP field and lies in a region of low background,
the offset between the closest $\bar{\chi ^2}$ minimum
and the association centroid was 8.4 $^{+1.8}_{-5.4}$ arcminutes
in the $-$x direction (roughly west), and 1.8 $\pm 1.8$ arcminutes
in the $+$y direction (roughly north).  The other minima showed
wide scatter about these values.

% For the brightest region in the WISP field, 
%  which coincided most closely with the 
% $\bar{\chi ^2}$ minima, the mean offset was 8.4 
% $^{+1.8}_{-5.4}$ arcminutes in the $-$x direction (roughly west),
% and 1.8 $\pm 1.8$ arcminutes in the $+$y direction (roughly
% north).  

The identifications, coordinates, and a brief description of
the 9 identified sources of scattered light are given in 
Table 2.  We also list the value of $\bar{\chi ^2}$ for
each object; these values are indicative of the degree to
which the polarization patterns near the objects resemble
the centro-symmetric halo that would be produced by scattering
of starlight from the central objects from a thin disk of material.

6 of the 9 regions are large OB associations
or complexes as identified by \cite{luc70}.  Two of the
others are large, very young open clusters; the last 
is an H {\small II} region associated with a small
starcloud (\cite{hen56}); the ionizing star is likely
to be the 11th magnitude B1 supergiant
Sk $-$70$\arcdeg$ 29 (\cite{san70}).  The illuminating 
sources in Table 2 are marked with star symbols
in Figure \ref{toon}. 

Due to our averaging procedure, and
because the errors in position angle do not form a
normal distribution, our calculated $\bar{\chi ^2}$ values
cannot be easily interpreted in absolute terms.  We therefore
do not place formal confidence limits on our detections of
scattering halos around the illuminating sources.  However,
the $\bar{\chi ^2}$ values listed in Table 2 do accurately 
reflect the {\it relative} strengths of the detected scattering
halos; the symbol sizes in Fig. \ref{toon} have been scaled
according to these values.

% If the errors in position angle
% are truly Gaussian, then our calculated $\chi ^2$ values
% can be directly translated into probabilities, which
% represent the confidence of our detection of scattering
% halos around the sources.  The probabilities thus estimated
% range from 10\% for N 11 to 1\% for the LH 16$+$17$+$20
% group; the symbol sizes in Fig. \ref{toon} are scaled 
% to these probability estimates.

In sharp contrast to LH 15 and the associations in N 11,
the group labelled B2 (notation from \cite{mar76}) corresponds
to a broad {\it maximum} in $\bar{\chi ^2}$, suggesting that the nearby
ISM does not support a scattering halo around the OB stars
within.  Similarly, the open cluster NGC 1818, which is of 
a similar age and brightness to the scattered light sources
NGC 1755 and NGC 1711, shows no evidence for a scattering halo.
These differences are discussed further in \S\ref{anal}.

\section{Analysis}
\label{anal}

\subsection{Correlations with other wavelengths}

The physical interpretation of our UV surface brightness and polarization
maps is greatly aided by the consideration of observations
at optical, infrared, and radio wavelengths.  UV polarization
maps probe the structure of the interstellar medium surrounding
young, massive stars.  Such extremely young stellar populations
are often attended by the signatures of recent star formation,
such as molecular gas, dark nebulae and emission from heated dust.
The dust which both absorbs and scatters the copious UV emission
from OB associations is distributed within the neutral hydrogen
clouds of the galactic disk.  While massive stars are the most
luminous, they are dwarfed in number by the less massive stars
which form along with them as well as those which have survived
from earlier star formation epochs; the combined light of low-mass stars
contributes to the UV surface brightness of the LMC, and will tend to
dilute the observed polarization.

Dust can be traced directly, via maps of the interstellar extinction
and by mapping the radiation from heated dust grains, and indirectly,
via maps of the distribution of \hi$ $ and estimates of the gas-to-dust
ratio.   In the infrared, it has been found that the 60 $\mu$m and 
100 $\mu$m data taken by the IRAS satellite correlate well with the
\hi$ $ distribution.
As discussed by \cite{sau91}, the bulk of the 60--100 $\mu$m emission
is due to the dust grains in the general ISM.  We note a good
correlation between the maps of \cite{sch89} and our UV image; 
the surface brightness at 60--100 $\mu$m and in our Fig. \ref{surf}
are of the same order of magnitude, with the IRAS values larger.
This confirms that Balmer continuum photons provide the bulk of 
the energy that heats the interstellar dust.

Maps of the interstellar reddening, e.g., \cite{iss84}, are
more sensitive to dust above the plane of the disk than within
it, but reveal that high amounts of extinction are present in 
front of N 11, in the area extending from LH 17 to LH 20, at 
the end of the stellar bar, and over a large region west of 
5$^{\mathrm h}$ right ascension and south of $-$67$\arcdeg$
30$\arcmin$.  There is little apparent correlation between
regions of high \ebv$ $ and the spiral structure studied by,
e.g., \cite{dev72}.  

\begin{figure*}
\centerline{\hbox{\psfig{figure=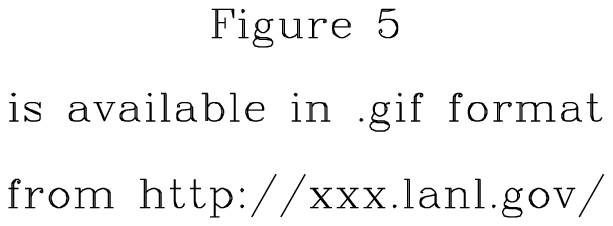}}}
\caption{A multiwavelength view of the western
LMC; {\sl far left:} 6600 \AA\ image from a 10 inch refractor (Sandage
1961); {\sl left center:} 2150 \AA\ image from WISP; 
{\sl right center:} 60 $\mu$m image from IRAS (Schwering 1989);
{\sl far right:} 21 cm H {\small I} image from the ATCA
(Kim {\it et al.} 1998).  The locations of supergiant H$\alpha$\
shells (Meaburn 1980) are overplotted on the optical image.
\label{tri}}
\end{figure*}

The distribution of \hi, mapped at low resolution by \cite{luk92}
and at high resolution by \cite{kim98}, correlates closely with
the IRAS maps; high-column density regions are associated with 
regions of high \ebv, although the correlation is not perfect.
These images are plotted for comparison in Figure \ref{tri}, 
along with an optical image.
For the direct UV light of OB associations, there is a clear
anti-correlation between \hi$ $ column density and UV brightness;
extreme examples are the areas of the supergiant H$\alpha$\ 
shells from \cite{mea80}. 
The shell LMC 1 surrounds the OB complex LH 15, and light from 
that association escapes virtually unabsorbed.  The same is true
for the giant shell around N 11; B2 is surrounded by a network of shells,
and LH 12 lies in the wall of the supergiant shell LMC 6.  The reason
for the anti-correlation 
is clear; stellar winds and supernovae clear the areas around 
prominent young associations and clusters.  However, when regions
far from the large associations are considered, the picture is
not as clear.  

We examined the maps of \cite{kim98}, looking for correlations
between structure in the \hi$ $ distribution and the UV light.
Figure \ref{tri} shows that instances can be found where the
diffuse UV light closely traces the \hi\ distribution.
Three of Meaburn's (1980) supergiant H$\alpha$\ shells
fall within the WISP field of view and are shown in the
lefthand panel of Figure \ref{tri}.
For example, the wall of LMC 1, which extends northeast
from N 11, shows a high UV surface brightness; throughout the
image, filaments and knots of \hi\ can be identified with 
UV emission.  The lack of \hi\ east-southeast of N 11 is 
seen as well in the UV; the same is true for the regions north
of NGC 1755, and in the extreme southwestern corner of the image. 
However, counterexamples
are present as well; the UV spur extending from LH 15 to NGC 1818
corresponds to a marked lack of \hi.

As shown in Figure \ref{tri}, the optical emission from the 
general stellar population provides a crucial piece of the 
puzzle as well.  Due to the smoother distribution of late-type
stars, the spiral arm structures are much less prominent in 
the optical image.  These stars, while faint in the UV, nevertheless
contribute by their sheer numbers to the UV surface brightness.
While individual stars are generally beneath our detection threshhold,
they contribute an unpolarized, unresolved background that will
tend to lower our polarization measures.  The stellar density is
highest in the southeast quadrant of our field, where we see the
western end of the LMC's bar.  The stellar density falls off rapidly
northward of $-$68$\arcdeg$, an impression confirmed 
by an examination of a starcount map produced from the USNO-A
astrometric catalog (Levine \& Monet, private communication).

It is apparent that we are looking at a field whose properties
show a strong north-south gradient.  In the south, the UV light
correlates well with general stellar density, identifications
with features in the \hi\ and IRAS maps are not obvious,
and potential sources of scattered light, such as B2, seem
less likely to contribute polarized flux to the UV light.  In 
the north, the field stellar density drops, the UV light 
correlates somewhat more strongly with the \hi\ surface 
brightness, and the large, young associations, such as N 11 
and LH 15, seem to support extended scattering halos.  

% $\bullet$ contains 11 dark nebulae from \cite{hod88}: no 
% real hope of identifying because of their scales (a couple
% pixels at most)...  and
% 10 molecular clouds from \cite{coh88}, 7 of which are
% north of about $-$67$\arcdeg$ 30$\arcmin$ $ $(some overlap
% between the two samples, but not extensive)...
% {\bf To Cohen's work, add \cite{is93}...}
% supergiant shells LMC 1, LMC 6, and LMC 7.  
% \cite{luk92} reported on the existence of an
% ``L-component'' to the \hi$ $ distribution that lies in
% front of the main disk; covers LH 19, 22, 21, 25, 12.

\subsection{Scattering geometry}

Given the general trends with environment explored in
the previous section, we now examine the effects of
the placement of illuminators and scatterers on the
observed polarization map.  We wish to uncover the 
similarities between the sources in Table 2;
to discover why the B2 complex and the cluster NGC 1818
seem not to contribute to the polarization map; and 
to assess the possibility that stars from beyond the WISP field
contribute to the scattered light. 

The most obvious requirement for the existence of scattering
halos around bright associations is the presence of interstellar
dust around the associations.  The asymmetry of the scattering
phase function must also be considered; there is strong evidence
for a forward-throwing UV phase function for interstellar dust
(e.g., \cite{mur95}, \cite{wit92}).  In these cases, back-scattering
is suppressed, and associations that lie between the observer
and a scattering layer will produce smaller scattered light
intensities than associations that lie behind the dust.

The WISP field of view encompasses two of the spiral arms
identified by \cite{dev72}: arm A extends from LH 15 down 
east side of our image to LH 17 and is defined by a chain
of clusters and associations; arm B extends from N 11
to the B2 complex in the south and is less well-defined 
in our image.  \cite{ste82} argued that spiral arms rich
in OB associations and H {\small II} regions should produce
more direct starlight than scattered light in the UV.  
Additionally, the lack of dust in the LMC's spiral arms 
(\cite{oes96}) will tend to suppress scattering.

A common feature of the sources in Table 2 that
support scattering halos is their placement within or 
at the edge of \hi$ $ giant or supergiant shells.  However,
given the generally chaotic distribution of \hi, this 
may be a difficult situation to avoid.  
 
One bright UV source that does not support a scattering
halo is NGC 1818, prominent in the UV spur containing LH 15.
NGC 1818 is 20--40 Myr old (\cite{wes97}), and is brighter
than both of the detected clusters NGC 1755 and NGC 1711
(V$_{1818}$ = 9.70, V$_{1755}$ = 9.85, V$_{1711}$ = 10.18;
\cite{van81}).
The region surrounding NGC 1818 is remarkably quiescent in
the \hi$ $ map, and the cluster itself is only moderately
reddened, with \ebv$ $ = 0.05.  The IRAS map (\cite{sch89})
also shows little thermal dust emission in the NGC 1818
region.  The WISP UV image shows moderate surface brightness
around NGC 1818, but in the higher resolution image of 
SCH, it begins to be resolved into individual stars.  
We conclude that NGC 1818 is prevented from contributing
to the polarization map by a lack of nearby dust.

The other perplexing association is the large B2 complex
in the southern WISP field.  Consisting of the Lucke-Hodge
associations LH 1, 2, 5, and 8, B2 is nearly the equal of N 11
in the WISP image, and appears to coincide with a high column
of \hi$ $.  The field stellar density around B2 is high, but
not as high as around the LH 16-17-20 area; these associations
are much smaller than B2 but still show some evidence of 
a scattering halo.  

There is clearly material available to scatter the starlight
of B2: its western half is highly reddened, an unambiguous
signature of dust.  The OB stars of B2 also power several
H {\small II} regions, most notably N 79 and N 83; these
regions are expected to appear as reflection nebulae in 
the UV.  We speculate in passing here that the LH 8 starcloud
may not be associated with the rest of B2; it is less highly
reddened and contains no large H {\small II} regions.
Moreover, it is spatially coincident with the LMC 7 supergiant
H$\alpha$\ shell of Meaburn (1980), corresponding to a 
hole in the \cite{kim98} \hi\ maps.
There is another
large hole in the \hi\ distribution at the 
western end of the bar (Kim, private communication).

One possibility is that the B2 complex is above the 
main disk of the LMC, and its scattering halo is 
therefore suppressed by the forward-scattering phase
function of the dust grains.  According to the channel
maps of \cite{kim98}, most of the \hi$ $ around B2 
lies at a heliocentric velocity of 241--274 km s$^{-1}$.
\cite{way90} has stated that interstellar absorption lines due to
LMC gas are present in stars in this region at velocities
of $\approx$220 km s$^{-1}$, but not at higher velocities,
implying that the illuminating stars lie above the main LMC disk.

A further possibility is that the large angular extent of
B2 exceeded the limits of the window we used to analyze 
$\bar{\chi^2}$ in the polarization map.  The 42 arcminute-square
window only just contained all of B2, and the point-source
approximation we used to test against the distribution of
position angles was almost certainly invalid for this case.
In addition, most of the window was probably swamped with
direct stellar light.
We repeated the $\bar{\chi ^2}$ analysis with
54 and 63 arcminute windows, but the results for B2 remained
unchanged.  The persistent non-detection leads us to suggest
that B2 does in fact lie above the main LMC disk, but this
is highly uncertain.

The results of our $\bar{\chi ^2}$ test suggest that
scattering of light from point sources by a smooth disk
cannot be the only source of polarization in the WISP
field.  In this case, it is possible that light is being
scattered from outside the WISP field of view, perhaps
from many kiloparsecs away, into our line of sight.
Prime candidates for illuminating sources would be the 
OB associations in Shapley Constellation IV (50$\arcmin$ 
east of LH 19), and the associations in the bar, extending 
southeast of LH 20.  Further afield, the 30 Dor association,
postulated by \cite{luk92} to lie as much as 400 pc above
the LMC midplane, could be a contributor to scattered light
in the WISP field.

To be able to scatter light from associations across the 
LMC, dust clouds must be raised above the plane of the 
disk, which is optically thick.  Among L$_{\star}$ spirals,
galaxies with comparable rates of star formation to the LMC have 
often had large amounts of dust ejected to distances
a kiloparsec or more above the plane (e.g., \cite{how97}).
Given the propensity of spiral galaxies for ejecting dust
into their halos, and the previously suggested depth
structure of the LMC, it is quite likely that the LMC 
possesses extraplanar dust; some of this dust may scatter
the light of distant OB associations.

\section{Discussion}
\label{dis}

\subsection{To What Extent is the Western LMC a Kiloparsec-Scale
Reflection Nebula?}

Several authors have suggested that UV-bright, gas rich galaxies
possess extended halos of scattered UV light, in effect creating
galaxy-sized reflection nebulae (e.g., \cite{ste82}, \cite{nef94}).
Characteristics of scattered light halos should include somewhat
bluer spectral energy distribution than the overall radiation field
of the galaxy, a correlation with the \hi$ $ distribution, and 
linear polarization.  Does the western side of the LMC disk
meet any of these criteria?

The WISP field is richer in young stars than the far outer
regions of spirals studied by Stecher {\it et al.} (1982) and
Neff {\it et al.} 1994.  Thus any putative scattered light
contribution to its UV surface brightness will be diluted by
direct, stellar light.  In the southern half of the WISP
field, the UV surface brightness correlates closely with 
optical images and with star count maps, suggesting that direct
light accounts for most of the ultraviolet radiation.
However, northward of $\approx$ $-$68$\arcdeg$, the stellar
density falls off rapidly; Figure \ref{tri} shows that the 
ultraviolet light falls off more slowly.  Throughout the
image, there are areas of diffuse UV emission that begin 
to be resolved in the images of SCH, but in the northern
WISP field it is easy to see correlations between diffuse
UV emission and the presence of \hi$ $ emission.  

The degree to which scattered light contributes to the 
total UV surface brightness of the LMC can be assessed
by examining the mean level of polarization.  Our error-weighted
mean P = 12.6\% $\pm$2.3\%; in galactic
reflection nebulae, the polarization at optical wavelengths
is typically $\approx$20--40\% (comparable to $p_{max}$;
e.g., \cite{wat91}).  We
assume that the pixels in which we find P = 30--40\% 
are analogous to reflection nebulae; light that has been
scattered nearly in the plane of the sky is the primary contributor
to their flux.  However, such high values are only found in 
areas of low flux, and thus high error, in our map. 

Because
the dust scale height of the LMC disk is non-negligible, and
because of the inclination of the disk to the plane of the sky,
typical polarization levels should fall below $p_{max}$.
Scarrott {\it et al.} (1990) showed the variation of P with
scattering angle for several dust grain size distributions.  
Polarization levels of 10--15\% are to be expected for scattering
angles $\theta$ $\approx$ 50$\arcdeg$--60$\arcdeg$ (and for
$\theta$ $\approx$ 125$\arcdeg$--135$\arcdeg$ if backscattering
is important).  These angles are close to the complement of the 
LMC's angle of inclination to the sky (28$\arcdeg$--45$\arcdeg$), 
and would be reasonable values for a mean angle of scattering.  

As noted in \S\ref{map}, most of our measured polarization
values lie between 5\% and 20\%; this implies that in the
average region of the WISP field, at least 20\%, and possibly
$>$90\% of the detected light has been scattered.  The statistical
significance of many of the polarization vectors is low; we seem
able to reliably detect polarizations where the contribution
of scattered light is $\gtrsim$50\%.

\cite{jur80} presented a method to predict the intensity
of scattered light from a dust cloud at a certain angular
distance from an illuminating source.  We adapted their
method to the case of forward-scattering grains; this 
model predicts that scattered light surface brightnesses
near complexes such as N11 and B2 should lie in the range
0.5 -- 1.5 \cgssb, within 30$\arcmin$ of the illuminating
sources.  This level should be taken as a strict lower limit,
since \cite{jur80} considered dust clouds in the plane of
the sky.   Accordingly,
the WISP data show surface brightnesses some 4--20 times 
higher than the \cite{jur80} formula predicts.  While some 
of this discrepancy is due to the integrated contribution
of scatterers along the line of sight (in front of or behind
the plane of the sky), it does suggest that scattered light
is not the only contributor to the total surface brightness.  The
largest difference between the predicted intensity due to
pure scattered light and the observed surface brightness 
occurs for the southern part of the WISP field, where the
field stellar density is high.

% \cite{jur80} presented a method to predict the intensity
% of scattered light from a dust cloud in an optically thin
% medium at a certain angular distance from an illuminating
% source.  We adapted their method to the case of forward-scattering
% dust grains; assuming values of $\tau$ between 0.3 and 1.3, 
% we estimate that the scattered light surface brightness near
% complexes such as N 11 and B2 should lie in the range 
% 0.5 -- 1.5 \cgssb, for distances within 30$\arcmin$ 
% from the associations.  Comparison to the WISP data 
% shows that these levels represent some 5--25\% of the 
% UV surface brightness around the associations.  The
% predicted levels of scattered light are a larger fraction
% of the total surface brightness for the associations
% in the northern part of the WISP image where the stellar
% density is low.

We conclude from the mean level of polarization and the
distributions of UV surface brightness around OB associations
that scattered light contributes at least 20\% to the total
UV surface brightness outside the associations, and possibly
more than 50\%.  The scattered
fraction is expected to be higher in the north, where stellar
densities are lower and we more confidently detect centro-symmetric
polarization patterns around the largest associations.  In this
respect, the northwestern disk of the LMC can be regarded as
a very faint UV reflection nebula.  The referee has
suggested that similar observations in another bandpass could
be a test of this hypothesis.  Because of the differing spectral
energy distributions of the reflected light of early-type stars
and the unresolved sea of late-type stars, an ultraviolet color 
map could test the idea that the scattered light fraction is
higher in the north.  Observations at 1500 \AA\ could be obtained,
although the ability to obtain sufficient signal-to-noise is
in question.

\subsection{What are the Scattering Geometry and Phase Function?}

The geometry of the scattering material is of crucial importance
for determining the distribution of polarizations.  We have seen
that the B2 complex shows no evidence for a scattering halo, and
suggested a possible geometric explanation.  Another issue is
the offset of centro-symmetric polarization patterns from the
central OB associations, noted in \S\ref{map}.  A possible 
explanation for this offset is the inclination of the LMC disk
to the line of sight.  The LMC is known to be tilted between 
28$\arcdeg$ and 45$\arcdeg$ from the plane of the sky, along 
a roughly north-south axis, with the eastern (30 Dor) side
closer to Earth (\cite{wes97}).  The effects of varying 
the inclination of a scattering disk have been investigated
by \cite{col99b} in this context, and this explanation does
seem plausible.

Another scattering issue that requires modelling is the
asymmetry in phase function: do the dust grains scatter 
isotropically, or are they strongly forward-throwing?  
If isotropic, the UV surface brightness should fall off
more slowly than for the forward-throwing case; the 
mean polarization level may also change as the preferred
paths for photon escape from the disk are changed.  For
isotropically scattering grains, the geometric explanation for 
B2's lack of scattering halo is weakened.

\section{Summary}
\label{sum}

We have obtained the first wide-field ultraviolet polarimetric
images in astronomy, using the rocket-borne Wide-Field Imaging
Survey Polarimeter (WISP).  We find the diffuse UV background
light is polarized at levels from 4--40\%, the lower limit
set by our sensitivity threshold.  

We photometrically calibrated our data using the data of \cite{smi87}
for the OB associations of \cite{luc70}; our photometric
accuracy is probably $\pm$ $\approx$20\%.
The UV surface brightness (Fig. \ref{cont})
is of the same order of magnitude as the IRAS 60$\mu$m and
100$\mu$m measures, indicating that Balmer continuum photons
are the primary heaters of interstellar dust in the LMC.  
We find a minimum UV surface brightness of 5.6 $\pm$3.1
\cgssb (23.6 $\pm$0.5 \surfb).  

A polarization map (Fig. \ref{pols}a) was created and analyzed in order to 
study the diffuse galactic light of the LMC.  The mean
level of polarization was P = 4.4\% $\pm$ 7.8\%; however,
we excluded regions where P $<$ 2.5$\sigma$ from the 
polarization map; the mean of the statistically significant
polarization vectors was P = 12.6\% $\pm$2.3\%.  This
suggests a mean contribution of scattered light to the
total of $>$20\%.   We used a modified $\chi ^2$ test
to identify the sources of scattered light within the 
WISP field, and found 9 bright regions that might be 
responsible for scattered light.  These are listed in 
Table 2 and Figure \ref{toon}.

The fall-off of surface brightness with radial distance
from the OB associations was examined and suggests that
the scattered light is not the sole contributor to the
observed surface brightness.
There is a north-south gradient of properties across the
field, due to the falloff in stellar density 
with distance from the bar. 
The northern half of the image contains the 
strongest evidence of centro-symmetric polarization patterns
and large amounts of scattered light contribute
to the UV surface brightness above $-$68$\arcdeg$.

We find that the B2 complex shows no evidence for a contribution
to the scattered light around it; it does not support a scattering
halo.  This may be an effect of the high field stellar density in the
neighborhood and unfavorable geometric positioning of stars 
relative to dust.  Alternatively, it may be the result of bias
introduced in our analysis of the polarization map.

Many of the questions raised by these data can only be 
answered at present through modelling efforts.  Some of
the issues of scattering geometry and phase function will
be addressed by \cite{col99b} in an accompanying paper.

\acknowledgments

WISP is supported by NASA grant NAG5--647.
This research has made use of the SIMBAD database of the Centre de 
Donn\'{e}es astronomiques de Strasbourg and NASA's Astrophysics
Data System Abstract Service.  We would like to thank the
anonymous referee whose comments improved the clarity of
this paper.

{}

\newpage

\end{document}